\documentstyle[psfig,conf_iap,]{article}
\def\hhhp{H$_{3}^{+}$}
\def\hh{H$_2$}
\def\hhp{H$_{2}^{+}$}
\begin{document}
\heading{Detection of \hhhp\ in the interstellar medium of
IRAS~08572+3915
} 
\par\medskip\noindent

\author{T. R. Geballe$^{1}$}

\address{
Gemini Observatory, 670 N. A'ohoku Place, Hilo, HI 96720, USA
}

\begin{abstract} 

The first detection of the molecular ion \hhhp\ in an extragalactic object
has been made toward the highly obscured ultraluminous galaxy
IRAS~08572+3915. Two absorption features in its spectrum near 3.9~$\mu$m
are identified as the redshifted 3.668~$\mu$m \hhhp\ doublet and
3.716~$\mu$m singlet lines of \hhhp, both previously detected in a number
of galactic dark and diffuse clouds. We discuss the probable location
of the \hhhp\ in this galaxy.

\end{abstract} 
 
\section{Introduction} 

\hhhp, the highly reactive molecular ion upon which interstellar gas phase
chemistry is based \cite{her73, wat73}, is important observationally for
understanding dark and diffuse interstellar clouds \cite{geb00}.  In both
types of clouds it is produced following cosmic ray ionization of \hh\ to
\hhp, which quickly reacts with \hh\ to form \hhhp.  \hhhp\ is destroyed
readily in dark clouds by reactions with neutral molecules (principally by
CO) and atoms (mainly by O) and even more readily in diffuse clouds by
dissociative recombination on electrons, which, due to the ionization of
carbon, are much more abundant than in dark clouds. The absorption
strengths of \hhhp\ lines provide basic information on the cloud
dimensions and environment, in addition to temperature.  Assuming a value
for the ionization rate, the column density of \hhhp\ directly yields
either the distance through the cloud to an embedded continuum source or
the thickness of an intervening cloud or group of clouds. Alternatively,
if an estimate for the line of sight distance through the absorbing
material is available, the ionization rate can be determined from the
column density.  These inferences are possible because unlike most other
molecules, the number density of \hhhp\ is a constant that depends only on
whether a cloud is dark or diffuse. This unusual property of \hhhp\ comes
about because its creation rate per unit volume is proportional to the
first power of the cloud density, rather than the square.

\hhhp\ has been detected in both dark and diffuse clouds by virtue of its
vibration-rotation transitions near 3.7~$\mu$m (an ortho-para doublet at
3.668~$\mu$m and a para singlet at 3.715~$\mu$m). The amounts of \hhhp\
that have been found in dark clouds are consistent with predictions based
on the estimated production rate and measured destruction rates
\cite{geb96, mcc99}. In diffuse clouds, however, there is a large
difference between the observed and predicted column densities
\cite{geb99, mcc02}; the cause may be the use of an inaccurate value for
the rate of dissociative recombination on electrons.

The detection of strong \hhhp\ absorption toward the center of the Galaxy
\cite{geb99} suggests that it is possible to detect \hhhp\ in the
interstellar medium of suitable external galaxies - those with
sufficiently bright and compact sources of infrared continuum radiation as
well as long columns of interstellar molecular gas along their lines of
sight. Although 8--10~m telescopes and their successors will be required
for studying \hhhp\ in all but a very few of the most brightest
candidates, at the United Kingdom 3.8~m Infrared Telescope (UKIRT) a
search was initiated in one of the most promising galaxies,
IRAS~08572+3915. This galaxy has a deep 10~$\mu$m silicate absorption
\cite{dud97} and the strongest 3.4~$\mu$m interstellar absorption feature
known \cite{wri96, ima00}, nearly twice as strong as the one observed
toward the Galactic center. The 3.4~$\mu$m feature is known to be
associated with objects reddened by diffuse gas \cite{aja99} and its
strength in IRAS~08572+3915 tends to imply the existence of a long
pathlength of molecular material along the line of sight to the nuclear
infrared source.

\begin{figure}
\centerline{\vbox{
\psfig{figure=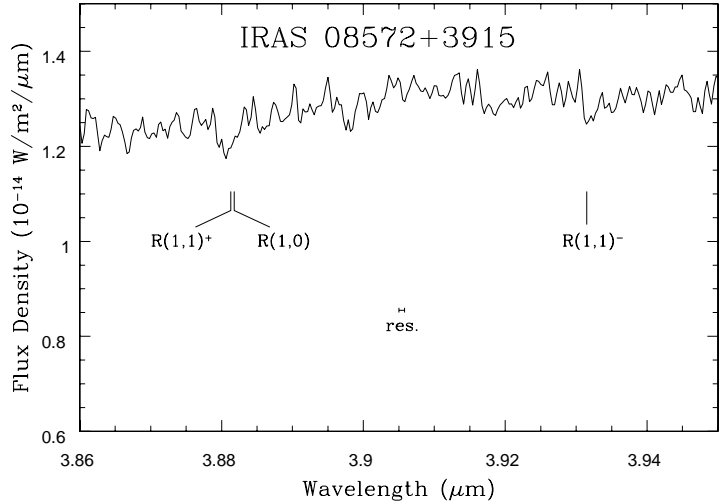,height=7.cm}
}}
\caption[]{The spectrum of IRAS~08572+3915 from 3.86 to 3.95~$\mu$m,
at R~$\sim$~5,000 The predicted wavelengths of \hhhp\ lines (using
z=0.05821; Solomon et al. 1997) are indicated.
}
\end{figure}

\section{Observations and results} 

In December 2000 a 3.82-3.98~$\mu$m spectrum of IRAS~08572+3915 was
obtained at UKIRT, using the facility spectrograph CGS4. The spectrum
covered the wavelength range in which the redshifted 3.7~$\mu$m lines of
\hhhp\ should appear. A portion of the spectrum is shown in Fig.~1 at a
resolving power of $\sim$~5,000. Statistically significant absorption
features are observed at the expected wavelengths of the two \hhhp\
features, 3.8815~$\mu$m and 3.9317~$\mu$m, indicating that the molecular
ion has been detected.

The 3.88~$\mu$m doublet, which is blended but which also would be
unresolved at the resolution used, appears to be roughly twice the
strength of the singlet, indicating that the doublet has roughly equal
contributions from its ortho and para components.  The column density of
\hhhp\ derived from the lines is about 5~$\times$~10$^{15}$~cm$^{-2}$,
roughly twice that found towards the Galactic center. This result is only
mildly sensitive to the temperature, which cannot be determined accurately
because of an insufficient signal-to-noise ratio and the low spectral
resolution, but which appears to be 100~K or less.  The FWHMs of the
absorptions cannot be determined accurately, but are probably
150$\pm$50~km~s$^{-1}$, similar to that found toward the Galactic center.

\section{Discussion and Conclusions}

The velocity at which \hhhp\ is seen in IRAS~08572+3915 is the systemic
velocity; hence the associated gas is neither falling into nor being
expelled from the presumed AGN \cite{dud97}. Presumably the \hhhp\ is
mainly found in relatively quiescent interstellar medium far from the
nucleus, as in the case of the Galactic center although in principle it
could be located in a circumnuclear molecular disk. Currently it is
uncertain whether the gas containing the detected \hhhp\ is in diffuse or
dark clouds.  However, the strong 3.4~$\mu$m interstellar feature suggests
that much of the gas must be diffuse. Assuming the canonical value of
$\sim$1~$\times$10$^{-7}$~cm$^{-3}$ for the density of \hhhp in diffuse
gas \cite{mcc98, geb99} the derived length of the column containing \hhhp\
in IRAS~08572+3915 is $\sim$~10~kpc, which seems excessive. A similar
unreasonably high result of several kpc was obtained for the Galactic
center \cite{geb99}. Using what is known by other means about the
dimensions of diffuse clouds in our galaxy, such as Cygnus OB2
\cite{geb99}, an \hhhp\ density in diffuse clouds is derived that is
roughly an order of magnitude higher than the canonical value, and for the
Galactic center and IRAS~08572+3915 column lengths equivalently lower, and
probably more sensible, are obtained.

An improved spectrum of \hhhp\ in IRAS~08572+3915 together with
observations and analysis of other molecular species are required to
better pinpoint the location of the \hhhp. New measurements of and
improved confidence in the rate coefficient for dissociative recombination
of \hhhp\ on electrons, which determines the steady-state abundance of
\hhhp\ in diffuse clouds, also is needed.  In the future, using
ground-based 8--10~m class and larger telescopes along with the NGST, one
can anticipate that spectroscopy of \hhhp, in combination with
measurements of dust, CO and other molecules, will be a standard technique
for probing the interstellar gas in many distant galaxies.

\acknowledgements{UKIRT is operated by the Joint Astronomy Centre on
behalf of the U.K. Particle Physics and Astronomy Research Council. I
thank the staff of UKIRT for its support. I am grateful to B. J. McCall
and T. Oka for helpful discussions.}

\vfill
\end{document}